\documentclass[proceedings]{JHEP37}
\usepackage{amsfonts}
\usepackage{amsmath}
\usepackage{epsfig}

\newbox\mybox

\newcommand\fverb{\setbox\mybox=\hbox\bgroup\verb}
\newcommand\fverbdo{\egroup\medskip\noindent\fbox{\unhbox\mybox}\ }
\newcommand\fverbit{\egroup\item[\fbox{\unhbox\mybox}]}
\conference{PT-symmetry breaking of nc-$E_2$ type}
\abstract{We propose a noncommutative version of the Euclidean Lie algebra $E_2$.
Several types of non-Hermitian Hamiltonian systems expressed in terms of generic
combinations of the generators of this algebra are investigated. Using the breakdown
of the explicitly constructed Dyson maps as a criterium, we identify the domains 
in the parameter space in which the 
Hamiltonians have real energy spectra and determine the exceptional points signifying the
crossover into the different types of spontaneously broken PT-symmetric regions 
with pairs of complex conjugate eigenvalues. We find exceptional points which remain invariant
under the deformation as well as exceptional points becoming dependent on the deformation
parameter of the algebra.}

\title{Spontaneous PT-symmetry breaking for systems of noncommutative
Euclidean Lie algebraic type}
\author{Sanjib Dey, Andreas Fring and Thilagarajah Mathanaranjan \\
Department of Mathematics, City University London,\\
Northampton Square, London EC1V 0HB, UK\\
E-mail: sanjib.dey.1@city.ac.uk, a.fring@city.ac.uk,
thilagarajah.mathanaranjan.1@city.ac.uk}

\input{tcilatex}
\begin{document}

\section{Introduction}

In \cite{DFM} we demonstrated that analogues of quasi-solvable models of Lie
algebraic type can be constructed in terms of Euclidean Lie algebra
generators. Unlike standard quasi-solvable models, this type of systems
admits solutions that can not be expressed in terms of hypergeometric
functions. Thus they constitute a different type of class than the more
common $sl_{2}(\mathbb{C})$-models \cite{Tur0} with their compact and
non-compact real forms $su(2)$ and $su(1,1)$ \cite{PEGAAF2,Paulos}. We
identified various types of $\mathcal{PT}$-symmetries for the $E_{2}$%
-algebra, which for concrete non-Hermitian models served to explain the
reality of their spectra in part of the parameter space. Similar features
were also previously observed for special cases of the $E_{2}$ \cite{BenKal}
and $E_{3}$ \cite{JonSmKal} Euclidean Lie algebra. Further interest in these
kind of models stems from the fact that for specific representations the
models become identical to some complex potential systems currently
investigated in optics \cite{Muss,MatMakris,Guo,OPMidya,MatHugh,MatHughEva}
and solid state physics \cite{MatLongo}.

Here we continue our investigations by considering deformations of the
systems studied in \cite{DFM}. In particular, we aim to identify the regions
in the parameter space where the models possess real eigenvalue spectra due
to intact $\mathcal{PT}$-symmetry and spontaneously broken $\mathcal{PT}$%
-symmetry where the eigenvalue spectra contain at least one pair of complex
conjugate eigenvalues, characterized by $[\mathcal{PT},H]=0$, $\mathcal{PT}%
\psi =\psi $ and $[\mathcal{PT},H]=0$, $\mathcal{PT}\psi \neq \psi $,
respectively \cite{EW,Bender:1998ke,Benderrev}. The transition from one to
the other region is marked by the so-called exceptional points \cite%
{Kato,HeissEx,IngridEx,IngridUwe}.

Alternatively, the two different regions are also characterized by a break
down of the so-called Dyson map $\eta $, defined as the map that adjointly
maps a non-Hermitian Hamiltonian $H\neq H^{\dagger }$ to isospectral
Hermitian counterparts $h=h^{\dagger }$ by means of $h=\eta H\eta ^{-1}$ 
\cite{Urubu,Alirev}. We will exploit here the latter criterium without
computing explicit solutions to the Schr\"{o}dinger equation, apart from
some exceptional cases.

\section{Deformations of the Euclidean E$_{2}$-Lie algebra and their $%
\mathcal{PT}$-symmetries}

We commence by introducing some natural deformations of the Euclidean Lie
algebra $E_{2}$, whose defining commutation relations for their three
generators $u$,$v$ and $j$ are 
\begin{equation}
\left[ u,j\right] =iv,\qquad \left[ v,j\right] =-iu,\qquad \text{and\qquad }%
\left[ u,v\right] =0.  \label{E2}
\end{equation}%
Some representations for this algebra may be found for instance in \cite{DFM}%
. A useful version for our purposes here, not reported in \cite{DFM}, is the
two-dimensional one%
\begin{equation}
j:=yp_{x}-xp_{y},\qquad u:=x,\qquad \text{and\qquad }v:=y,  \label{rep2}
\end{equation}%
expressed in terms of generators of the Heisenberg canonical variables $%
x,y,p_{x},p_{y}$ with non-vanishing commutators $\left[ x,p_{x}\right] =%
\left[ y,p_{y}\right] =i$. We have set the reduced Planck constant to $\hbar
=1$. We can now simply consider various two-dimensional canonical spaces and
investigate how they are translated into the $E_{2}$-setting. The most
evident choice would be to replace the canonical variables in (\ref{rep2})
with some variable on a flat noncommutative space. In general, that does not
lead to a closure of the algebra. However, when we carry out a Bopp-shift
only in the $u$ and $v$ generators, we obtain the new generators%
\begin{equation}
J:=yp_{x}-xp_{y},\qquad U:=x-\frac{\theta }{2}p_{y},\qquad \text{and\qquad }%
V:=y+\frac{\theta }{2}p_{x},  \label{repE2}
\end{equation}%
obeying the deformed $E_{2}$-algebra%
\begin{equation}
\left[ U,J\right] =iV,\qquad \left[ V,J\right] =-iU,\qquad \text{and\qquad }%
\left[ U,V\right] =i\theta .  \label{NCE2}
\end{equation}%
We notice that the deformed $E_{2}$-algebra is left invariant with regard to
following antilinear maps \cite{EW} reported for the $E_{2}$-algebra in \cite%
{DFM} 
\begin{equation}
\begin{array}{lllll}
\mathcal{PT}_{3}: & J\rightarrow J, & U\rightarrow V, & V\rightarrow U, & 
i\rightarrow -i, \\ 
\mathcal{PT}_{4}: & J\rightarrow J, & U\rightarrow -U, & V\rightarrow V, & 
i\rightarrow -i, \\ 
\mathcal{PT}_{5}: & J\rightarrow J, & U\rightarrow U, & V\rightarrow -V, & 
i\rightarrow -i.%
\end{array}
\label{PT}
\end{equation}%
We also observe that the additional symmetries, which hold for the original $%
E_{2}$-algebra (\ref{E2}), $\mathcal{PT}_{1}:$ $j\rightarrow -j$, $%
u\rightarrow -u$, $v\rightarrow -v$, $i\rightarrow -i~$and $\mathcal{PT}%
_{2}: $ $j\rightarrow -j$, $u\rightarrow u$, $v\rightarrow v$, $i\rightarrow
-i$, are broken for the deformed version (\ref{NCE2}). While these
symmetries are completely excluded here, it was found in \cite{DFM} that
they lead to less interesting models in the undeformed case as the
considered general Hamiltonians in the $E_{2}$-generators become Hermitian
without any further transformation needed by implementing the constraints
discussed therein.

\section{$\mathcal{PT}$-symmetric and spontaneously broken regions in
parameter space}

Let us now indicate how we distinguish the $\mathcal{PT}$-symmetric regions
in parameter space from the spontaneously broken $\mathcal{PT}$-symmetric
ones by analyzing the properties of the Dyson map. We start from
non-Hermitian $\mathcal{PT}_{3/4/5}$-invariant Hamiltonians $H$ in term of
generators of the deformed $E_{2}$-algebra (\ref{NCE2}). We consider all
Hamiltonians of this type that may be brought into the general form 
\begin{equation}
H_{\mathcal{PT}}(U,V,J)=\hat{c}_{1}J^{2}+\hat{c}_{2}J+\hat{c}_{3}U+\hat{c}%
_{4}V+\hat{c}_{5}U\!J+\hat{c}_{6}VJ+\hat{c}_{7}U^{2}+\hat{c}_{8}V^{2}+\hat{c}%
_{9}UV+\hat{c}_{10}  \label{HPT}
\end{equation}%
with $\hat{c}_{j}=\hat{\alpha}_{j}+i\hat{\beta}_{j}$ for $j=1,\ldots ,10$
and $\hat{\alpha}_{j},\hat{\beta}_{j}\in \mathbb{R}$. The specific form of
the constants $\hat{c}_{j}$ of being either real or purely imaginary is
governed by the particular $\mathcal{PT}$-symmetry we wish to implement.
Similarly as in \cite{DFM} we act adjointly with the so-called Dyson map on
the Hamiltonian in (\ref{HPT}) and demand that the resulting expression is
Hermitian, that is $\eta H\eta ^{-1}=h=h^{\dagger }$. The transformed
Hamiltonian will be of the same general form as $H_{\mathcal{PT}}$ 
\begin{equation}
h_{\mathcal{PT}}(U,V,J)=c_{1}J^{2}+c_{2}J+c_{3}U+c_{4}V+c_{5}U%
\!J+c_{6}VJ+c_{7}U^{2}+c_{8}V^{2}+c_{9}UV+c_{10},  \label{hkl}
\end{equation}%
albeit with different constants $c_{j}=\alpha _{j}+i\beta _{j}$ for $%
j=1,\ldots ,10$ and $\alpha _{j},\beta _{j}\in \mathbb{R}$. Computing then $%
h_{\mathcal{PT}}=h_{\mathcal{PT}}^{\dagger }$ \ for (\ref{hkl}) leads to the
following ten constraints%
\begin{equation}
\begin{array}{lllll}
\beta _{1}=0,~~~ & \beta _{2}=0,~~~ & \beta _{3}=\frac{\alpha _{6}}{2},~~~ & 
\beta _{4}=-\frac{\alpha _{5}}{2},~~~ & \beta _{5}=0, \\ 
\beta _{6}=0,~~ & \beta _{7}=0,~~ & \beta _{8}=0,~~ & \beta _{9}=0,~~ & 
\beta _{10}=-\frac{\theta \alpha _{9}}{2},~~%
\end{array}
\label{constraint}
\end{equation}%
with all remaining constants left to be unrestricted. Thus provided we can
compute the transformation from $H_{\mathcal{PT}}$ to $h_{\mathcal{PT}}$ in
a well-defined manner, the solution of the ten equations (\ref{constraint})
will characterize the domain in the parameter space, spanned by the
constants $\hat{c}_{j}$, for which the eigenvalues of $H_{\mathcal{PT}}$ are
guaranteed to be real.

We assume here for this purpose the Dyson map to be linear in the deformed $%
E_{2}$-generators in the exponential 
\begin{equation}
\eta =e^{\lambda J+\rho U+\tau V},\qquad \text{\ \ \ \ \ \ \ \ for }\lambda
,\tau ,\rho \in \mathbb{R},  \label{eta}
\end{equation}%
such that we can easily compute the adjoint action of this operator on the
deformed $E_{2}$-generators. We obtain%
\begin{eqnarray}
\eta J\eta ^{-1} &=&J+i(\rho V-\tau U)\frac{\sinh \lambda }{\lambda }+\left[
\rho U+\tau V+\frac{\theta }{\lambda }(\rho ^{2}+\tau ^{2})\right] \frac{%
1-\cosh \lambda }{\lambda },  \label{ad1} \\
\eta U\eta ^{-1} &=&\left( U+\frac{\rho \theta }{\lambda }\right) \cosh
\lambda -i\left( V+\frac{\tau \theta }{\lambda }\right) \sinh \lambda -\frac{%
\rho \theta }{\lambda },  \label{ad2} \\
\eta V\eta ^{-1} &=&\left( V+\frac{\tau \theta }{\lambda }\right) \cosh
\lambda +i\left( U+\frac{\rho \theta }{\lambda }\right) \sinh \lambda -\frac{%
\tau \theta }{\lambda }.  \label{ad3}
\end{eqnarray}%
As expected, we recover the expressions for the undeformed $E_{2}$-Lie
algebraic reported in \cite{DFM} in the limit $\theta \rightarrow 0$.

Notice that for Hermitian representations of $J$, $U\,$\ and $V$ also $\eta $
will also be Hermitian. Thus whenever the assumptions of this transformation
are respected the eigenvalues of $H_{\mathcal{PT}}$ must be real by
construction. This means that when the constraints (\ref{constraint}) allow $%
\lambda ,\tau ,\rho \in \mathbb{R}$ we are in the $\mathcal{PT}$-symmetric
regions in parameter space where $H_{\mathcal{PT}}$ has real eigenvalues. In
turn when the equations (\ref{constraint}) impose $\lambda ,\tau ,\rho
\notin \mathbb{R}$ the $\mathcal{PT}$-symmetry is spontaneously broken and $%
H_{\mathcal{PT}}$ acquires complex eigenvalues in form of conjugate pairs.

\subsection{ $\mathcal{PT}_{5}$-symmetric Hamiltonians of deformed $E_{2}$
Lie algebraic type}

We present here a detailed analysis of the $\mathcal{PT}_{5}$-symmetric
non-Hermitian Hamiltonians, because these were also the models worked out in
most detail for the undeformed version in \cite{DFM}. The $\mathcal{PT}_{5}$
invariant Hamiltonian 
\begin{equation}
H_{\mathcal{PT}_{5}}(U,V,J)=\mu _{1}J^{2}+\mu _{2}J+\mu _{3}U+i\mu _{4}V+\mu
_{5}U~\!\!J+i\mu _{6}V~\!\!J+\mu _{7}U^{2}+\mu _{8}V^{2}+i\mu _{9}UV.~~
\end{equation}%
is non-Hermitian unless $\mu _{6}=\mu _{5}+2\mu _{4}=0$. In the undeformed
case we found that the minimal requirement to satisfy the constraints (\ref%
{constraint}) is 
\begin{equation}
\tau =0,\quad \rho =\frac{\lambda \left( \mu _{5}-\mu _{6}\coth \lambda
\right) }{2\mu _{1}},\quad \coth (2\lambda )=\frac{\mu _{78}}{\mu _{19}}%
,\quad \coth \lambda =\frac{\mu _{23}}{\mu _{24}}.  \label{c2}
\end{equation}%
For convenience we introduced here the abbreviations%
\begin{equation*}
\mu _{78}:=\frac{\mu _{5}^{2}+\mu _{6}^{2}}{4\mu _{1}}-\mu _{7}+\mu
_{8},~~\mu _{19}:=\frac{\mu _{5}\mu _{6}}{2\mu _{1}}-\mu _{9},~~\mu _{23}:=%
\frac{\mu _{2}\mu _{5}}{2\mu _{1}}-\frac{\mu _{6}}{2}-\mu _{3},~~\mu _{24}:=%
\frac{\mu _{2}\mu _{6}}{2\mu _{1}}-\frac{\mu _{5}}{2}-\mu _{4}.
\end{equation*}%
Thus, according to the last two constraints in (\ref{c2}), the domain in the
parameter space for which $H_{\mathcal{PT}_{5}}(U,V,J)$ is guaranteed to
have real eigenvalues is characterized by the two inequalities%
\begin{equation}
\left\vert \mu _{78}\right\vert \geq \left\vert \mu _{19}\right\vert \text{%
\quad \quad and\quad \quad }\left\vert \mu _{23}\right\vert \geq \left\vert
\mu _{24}\right\vert .  \label{2un}
\end{equation}%
Surprisingly when we carry out the analysis for the deformed algebra the
first three constraints in (\ref{c2}) remain completely unchanged. The last
one is replaced by 
\begin{equation}
\mu _{3}=\mu _{3}^{\theta =0}+\theta 2\mu _{56}\left[ \mu _{19}\left( \frac{1%
}{2}+\cosh \lambda \right) -\mu _{78}\sinh \lambda -\mu _{68}\frac{1+\cosh
\lambda }{\sinh \lambda }\right] ,  \label{mu3}
\end{equation}%
where we introduced%
\begin{equation}
\mu _{56}:=\frac{\mu _{6}\cosh \lambda -\mu _{5}\sinh \lambda }{2\mu
_{1}(1+\cosh \lambda )}\qquad \text{and\qquad }\mu _{68}:=\frac{\mu _{6}^{2}%
}{4\mu _{1}}+\mu _{8}.
\end{equation}%
For brevity we have also denoted here by $\mu _{3}^{\theta =0}=-\mu
_{24}\coth \lambda +\mu _{2}\mu _{5}/(2\mu _{1})-\mu _{6}/2$ the value of $%
\mu _{3}$ as reported in \cite{DFM}, i.e. the value obtained from solving
from the last constraint in (\ref{c2}) for $\mu _{3}$. This means the
exceptional points resulting from the violation of the first inequality in (%
\ref{2un}) remain invariant under the deformation, whereas the one resulting
from the second inequality acquires a $\theta $-dependence. It is clear from
(\ref{mu3}) that a generic discussion would be rather involved. Since the
invariant exceptional point is less interesting, let us therefore make a
special choice $\mu _{7}=(\mu _{5}^{2}+\mu _{6}^{2}+4\mu _{1}\mu _{8})/(4\mu
_{1})$ and $\mu _{9}=\mu _{5}\mu _{6}/(2\mu _{1})$ for which the third
constraint in (\ref{c2}) no longer emerges. With this choice, equation (\ref%
{mu3}) can be brought into the form%
\begin{equation}
\coth \lambda =\frac{\mu _{1}\mu _{23}+\theta \mu _{5}\mu _{68}}{\mu _{1}\mu
_{24}+\theta \mu _{6}\mu _{68}}.  \label{cl}
\end{equation}%
The isospectral Hermitian Hamiltonian resulting from the similarity
transformation with these constraints is then computed to%
\begin{eqnarray}
h_{\mathcal{PT}_{5}}(U,V,J) &=&\mu _{1}J^{2}+\left( \mu _{2}+\theta \mu
_{6}\mu _{56}\right) J+\left( \mu _{8}+\frac{\mu _{5}^{2}}{4\mu _{1}}+\mu
_{6}\mu _{56}\right) U^{2}+\mu _{68}V^{2} \\
&&+\left[ \frac{\mu _{2}}{\mu _{1}}\mu _{65}-\mu _{23}\cosh \lambda +\mu
_{24}\sinh \lambda +\theta \mu _{56}\left( \frac{\mu _{6}}{\mu _{1}}\mu
_{65}+2\mu _{68}\sinh \lambda \right) \right] ~U  \notag \\
&&+\mu _{65}\{U,J\}-\theta \left[ \frac{\mu _{5}\mu _{6}}{4\mu _{1}}+\mu
_{56}\left( \mu _{24}\cosh \lambda -\mu _{23}\sinh \lambda -\mu _{4}-\frac{%
\mu _{5}}{2}\right) \right]   \notag \\
&&-\theta ^{2}\mu _{56}^{2}\left[ \mu _{68}(1+2\cosh \lambda )+\mu _{8}%
\right] ,  \notag
\end{eqnarray}%
where as usual $\{A,B\}:=AB+BA$ denotes the anti-commutator and%
\begin{equation}
\mu _{65}:=\frac{\mu _{5}}{2}-\frac{\mu _{6}}{2}\tanh \frac{\lambda }{2}.
\end{equation}%
By construction the eigenvalues of this Hamiltonian are real when the
absolute value of the right hand side of (\ref{cl}) is greater than 1.

Making now the additional parameter choice%
\begin{equation}
\mu _{2}=0,\quad \mu _{5}=-2\mu _{4},\quad \mu _{6}=-2\mu _{3},\quad \mu
_{8}=-\frac{\mu _{3}^{2}}{\mu _{1}},\qquad
\end{equation}%
also the constraint (\ref{c2}) vanishes. This means the Dyson map is a
well-defined transformation for all choices of the remaining free
parameters. Thus we do not expect any $\mathcal{PT}$-symmetry breaking and
all eigenvalues to be real. Indeed this is easily verified. Choosing
furthermore $\mu _{3}=\mu _{4}\coth (\lambda /2)$, the isospectral Hermitian
Hamiltonian simply results to%
\begin{equation}
h_{\mathcal{PT}_{5}}(U,V,J)=\mu _{1}J^{2}+\varepsilon J+\frac{\theta \mu
_{4}^{2}}{\mu _{1}}\left[ \varepsilon -\coth \left( \frac{\lambda }{2}%
\right) \right] ,  \label{sim}
\end{equation}%
with $\varepsilon =\theta \mu _{4}^{2}/\left[ \mu _{1}\sinh ^{2}(\lambda /2)%
\right] $. Using the representation (\ref{repE2}) in coordinate space with $%
p_{x,y}=-i\hbar \partial _{x,y}$, the Schr\"{o}dinger equation $h_{\mathcal{%
PT}_{5}}\psi =E\psi $ converts into%
\begin{equation}
-\mu _{1}\partial _{\varphi }^{2}\psi (\varphi ,r)+i\varepsilon \partial
_{\varphi }\psi (\varphi ,r)=E\psi (\varphi ,r),  \label{eq}
\end{equation}%
when using polar coordinates $x=r\cos \varphi $ and $y=r\sin \varphi $.
Equation (\ref{eq}) is easily solved to%
\begin{equation}
\psi (\varphi ,r)=c_{1}(r)e^{i\left( \varepsilon +\sqrt{\varepsilon
^{2}+4E\mu _{1}}\right) /(2\mu _{1})\varphi }+c_{2}(r)e^{i\left( \varepsilon
-\sqrt{\varepsilon ^{2}+4E\mu _{1}}\right) /(2\mu _{1})\varphi },
\end{equation}%
with $c_{1}(r)$, $c_{2}(r)$ being generic functions of $r$ and arbitrary
energies $E$. Quantizing now the system by demanding the periodicity $\psi
(\varphi +2\pi ,r)=\psi (\varphi ,r)$ the energy discretises to%
\begin{equation}
E_{n}=4\pi ^{2}\mu _{1}n^{2}-\theta \frac{2\pi \mu _{4}^{2}}{\mu _{1}\sinh
^{2}(\lambda /2)}n.
\end{equation}%
As expected this energy is always real and no $\mathcal{PT}$-symmetry
breaking can occur.

We have carried out a similar analysis for $\mathcal{PT}_{3}$ and $\mathcal{%
PT}_{4}$-symmetric systems and found identical qualitative behaviour in the
sense that the exceptional points governed by one of the inequalities remain
invariant under the transformation and those characterized by a second
inequality acquires a $\theta $-dependence.

\section{Conclusion}

We have introduced a natural and self-consistent deformation of the
Euclidean Lie algebra $E_{2}$, by studying the effects of Bopp-shifts in
some explicit representations. Of the previously \cite{DFM} identified five
different types of antilinear, e.g. $\mathcal{PT}$, symmetries for this
algebra two of them were found to be broken by the deformation. For the
remaining ones we studied invariant non-Hermitian systems expressed in terms
of the generators of the deformed algebra (\ref{NCE2}). The main question
addressed in this manuscript was to identify the domains in the parameter
space where the non-Hermitian Hamiltonians possess real eigenvalues. This
question was answered without the explicit computation of the eigenvalue
spectrum by identifying instead the regions for which the Dyson map breaks
down, which concretely meant that we derived a simply criterium for which
the parameter $\lambda $ in the transformation ceases to be real and becomes
complex. The transition is characterized by the so-called exceptional
points. We found two qualitatively different types of behaviour, one in
which the exceptional points remain invariant under the deformation and one
in which the corresponding energies acquire an explicit dependence on the
deformation parameter $\theta $. For the $\mathcal{PT}_{5}$-symmetric
Hamiltonian we provided a detailed analysis by successively fixing more and
more of the free parameters, thus passing through the stages of having
initially all types of exceptional points in the model, then having only the
non-invariant ones and finally for a choice that eliminates all of the
exceptional points and therefore excluding all possibilities of $\mathcal{PT}
$-symmetry breaking.

Evidently there are many interesting open challenges left. Naturally one
might investigate different types of deformations of the Euclidean Lie
algebra $E_{2}$ and also extend the analysis to higher ranks. In regard to
the physical questions addressed in this context the precise nature of the
breaking of the different $\mathcal{PT}$-symmetries would be interesting to
investigate. Except for the simple model (\ref{sim}), we have circumvented
here the explicit construction of the eigenvalue systems, but it would of
course be valuable to establish whether the models considered are indeed
solvable and confirm the spontaneous $\mathcal{PT}$-symmetry breaking also
on the basis of the explicit eigenvalues.

\noindent \textbf{Acknowledgments:} SD is supported by a City University
Research Fellowship. TM is funded by an Erasmus Mundus scholarship and
thanks City University for kind hospitality.


\end{document}